\documentclass[aip,pop,reprint,floatfix]{revtex4-1}
\usepackage{graphicx}
\usepackage[breaklinks=true,colorlinks=true,linkcolor=blue,urlcolor=blue,citecolor=blue]{hyperref}

\newsavebox{\myimage}
\newcommand\redsout{\bgroup\markoverwith{\textcolor{red}{\rule[0.5ex]{2pt}{0.4pt}}}\ULon}

\draft 

\begin{document}

\title{On the hysteresis in fireball formation and extinction}

\author{Brett Scheiner}
\email[]{bss@lanl.gov}
\affiliation{Computational Physics Division, Los Alamos National Laboratory, Los Alamos NM 87545  }

\author{Lucas Beving}
\affiliation{Department of Physics and Astronomy, University of Iowa, Iowa City, IA 52240}

\author{Scott D. Baalrud}
\affiliation{Department of Physics and Astronomy, University of Iowa, Iowa City, IA 52240}

\date{\today}

\begin{abstract}

A model is proposed to explain hysteresis observed in fireball formation and extinction as electrode bias is varied in partially ionized plasmas. Formation is predicted after a sufficiently deep potential well is established in the electron sheath of the electrode. Under the experimental conditions considered, once the fireball forms the plasma potential rapidly increases, resulting in electrons being only lost to the electrode.Previous predictions suggest that once formed the fireball double layer must maintain a potential close to the ionization potential of the neutral gas to remain in a steady state. In this paper, it is predicted that changes in electrode bias after formation results in a corresponding change in fireball size and plasma potential. This change in plasma potential allows the double layer potential to be maintained at biases both above and below the electrode bias at onset. The fireball extinguishes when the required double layer potential can no longer be maintained with balance of current loss of the bulk plasma. These predictions are tested experimentally and are found to be in good agreement with the measurements.

\end{abstract}


\maketitle

Fireballs are a discharge phenomenon that can occur near electrodes biased above the plasma potential by an amount greater than the neutral gas ionization potential\cite{1929PhRv...33..954L}. They are characterized by a secondary plasma whose plasma potential is near the electrode potential and at its boundary is separated from the bulk plasma by a double layer. 
Typically, the fireball plasma is several hundred Debye lengths, much larger than the initial sheath scale. Recent experimental studies of fireballs have focused on their stability\cite{2011PhPl...18f2113S,2013PSST...22f5002Y,2008PSST...17c5006S,2012PSST...21a5012S} and properties as ion sources\cite{2017RScI...88f3507C,2011RScI...82l3303P,2014PhPl...21b3505M}. 
One of the most common observations is hysteresis in the current-voltage (I-V) traces between the upswing and downswing of the electrode bias\cite{1991JPhD...24.1789S,2009PSST...18c5002B}, see Fig. 1A. Upon increasing the electrode bias, the fireball onset is abrupt lasting on the order of a microsecond once a critical electrode bias is exceeded\cite{1991JPhD...24.1789S,doi:10.1063/1.5026869}. This critical bias is determined by the neutral gas pressure, electrode area, and the electron-impact ionization cross section of the neutral gas\cite{2017PhPl...24k3520S}.
After onset, increased current collection is observed primarily due to the greater electron collection by the surface area of the fireball compared to that of the initial electron sheath, but also due to an increased ionization rate within the fireball. Upon decrementing the electrode bias below the critical bias, the fireball persists and the increased current collection relative to the pre-fireball formation value continues to be observed. The fireball eventually collapses after continued decrementation of the electrode bias and the electron current collection returns to pre-fireball levels. 

When the plasma chamber is small enough, fireball formation result in a state of global non-ambipolar flow where electrons are only lost to the fireball surface area\cite{2007PhPl...14d2109B}. In this state, the bulk plasma potential is locked to a fixed offset of the potential of the electrode due to the nearly exclusive collection of electrons by the electrode resulting from the relatively large ratio of electron collection area, represented by the fireball surface area, to ion collection area represented by the plasma chamber wall. This behavior has been observed in previous experimental studies\cite{2009PSST...18c5002B,1991JPhD...24.1789S}. The description in this paper concerns plasmas in which changes in the fireball size affect the bulk plasma potential where, as to be shown, the fireball area is limited by the chamber size and the requirement for global current balance. 


In this paper, a mechanism for this hysteresis is presented. The critical electrode bias required for fireball onset was described in a previous work as being due to formation of an ion rich layer by increased ionization in the electron sheath\cite{2017PhPl...24k3520S}. Here, the model is extended to predict the minimum electrode bias that can maintain a fireball after formation. 
Fireballs in steady state have been observed and theoretically predicted to have a double layer potential near the ionization potential of the neutral gas\cite{1991JPhD...24.1789S,2009PSST...18c5002B,2017PhPl...24k3520S}. A double layer potential near the ionization potential is needed to preserve particle and power balance of the fireball\cite{2017PhPl...24k3520S}. This paper predicts that as the electrode bias varies the surface area of the fireball will change, causing changes in the plasma potential and allowing the required double layer potential to be maintained. 
The fireball is predicted to extinguish only at a sufficiently low electrode bias that the minimum double layer potential can no longer be maintained. These predictions are shown to compare well with new experimental measurements. 


\begin{figure}
\includegraphics[scale=.75]{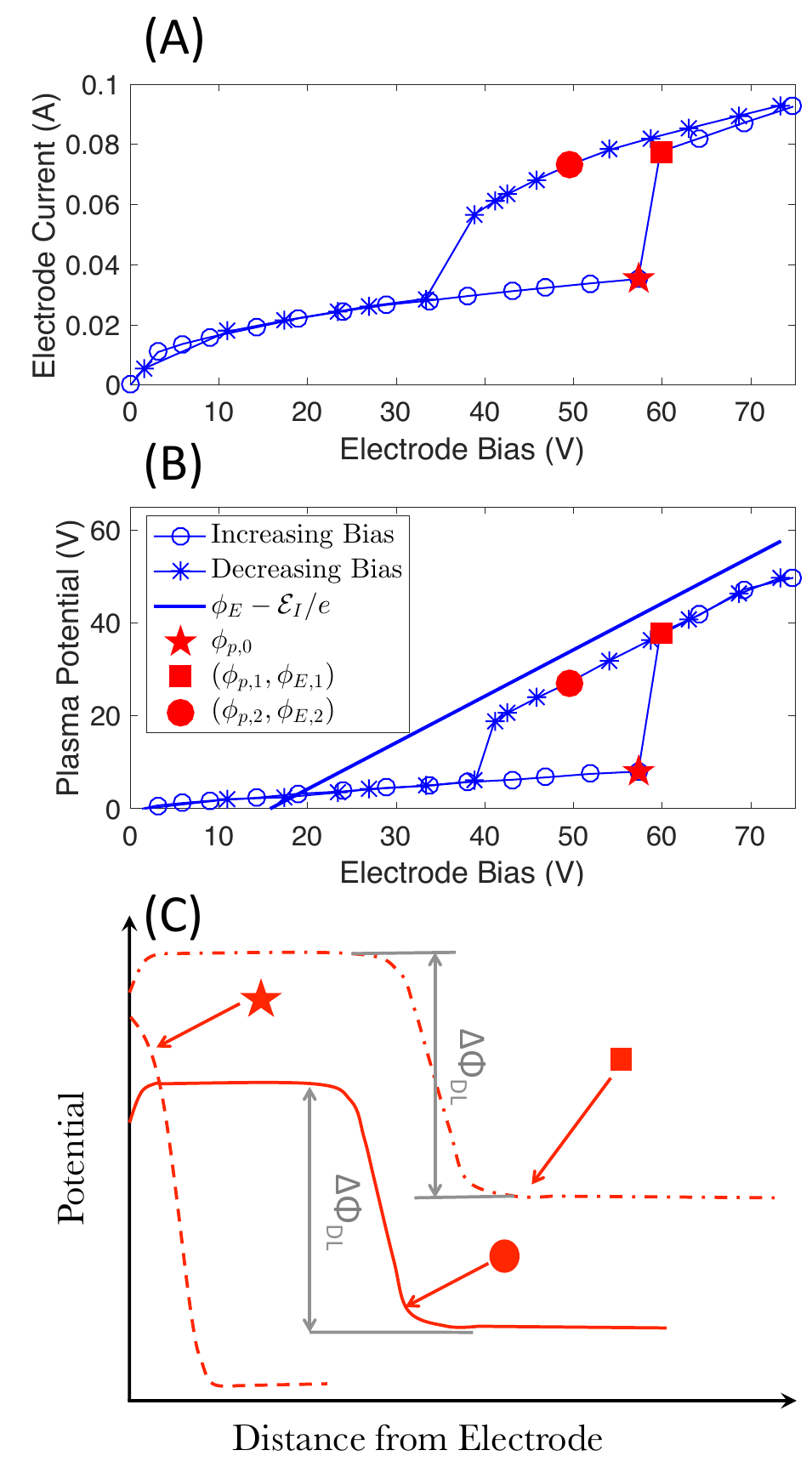}
\caption{A) The measured electrode I-V characteristic exhibiting hysteresis with red markers indicating different features of the hysteresis described in the text. B) The corresponding plasma potential values. C) Cartoon potential profiles typical of those which correspond to the conditions indicated by the red markers in A) and B). }
\end{figure}

In the model of Ref. \onlinecite{2017PhPl...24k3520S}, the critical electrode bias for fireball onset relative to the plasma potential, $\Delta\phi_c$, was determined by requiring that the ion birth rate due to electron-impact ionization of neutrals within the electron sheath exceeds the ion loss rate to the bulk plasma through the sheath surface. Once this occurs, a potential well forms due to the buildup of ions in the sheath, which traps low energy electrons born from ionization and forms a quasineutral fireball plasma. In the model, the critical electrode bias is determined by
\begin{equation}\label{tc2}
\frac{1}{2}\frac{A_{\text{S}}}{A_{\text{E}}}\sqrt{\frac{m_i}{m_e}}n_n\sigma_I(e\Delta\phi_c)z_{\text{S}}(e\Delta\phi_c)=1,
\end{equation}
where 
\begin{equation}\label{esh}
z_{\text{S}}(e\Delta\phi_c)=0.79\lambda_{De}\bigg(\frac{e\Delta\phi_c}{T_e}\bigg)^{3/4} 
\end{equation}
is the electron sheath thickness\cite{2015PhPl...22l3520S,2005PhPl...12e5502H} and 
\begin{equation}
\frac{A_{\text{S}}}{A_{\text{E}}}=\frac{\pi D z_{\text{S}}+\pi(D/2)^2}{\pi(D/2)^2}
\end{equation}
is the ratio of sheath surface area to electrode surface area. Given data for the energy dependent electron-impact ionization cross section of the neutral species and its density, $\sigma_I(\mathcal{E}=e\Delta\phi_c)$ and $n_n$, the critical bias can be solved. This produced estimates for the critical bias which accurately predicted experimental values for variation in both pressure and electrode size observed in Ref.~\onlinecite{2009PSST...18c5002B}.


The model of Ref.~\onlinecite{2017PhPl...24k3520S} also predicted requirements for a steady-state fireball by considering particle and power balance of the fireball plasma along with the flux balance given by the Langmuir condition $\Gamma_e=\Gamma_i\sqrt{m_i/m_e}$, where $\Gamma_{e,i}$ are the flux densities of the electrons and ions at the boundary of the double layer. 
The condition for the double layer potential was found to be
\begin{equation}\label{pdli}
e\Delta\phi_{DL}=\mathcal{E}_I+\mathcal{O}(T_{e,\text{F}}),
\end{equation}
where $\mathcal{O}(T_{e,\text{F}})$ is a term that is of the order of the electron temperature in the fireball, typically a few eV. A value of $e\Delta\phi_{DL}$ greater than that given by Eq.~\ref{pdli} results in an imbalance in the steady-state Langmuir condition due to an overdensity in the fireball. Such an overdensity leads to outward motion of the double layer surface
\footnote{The Langmuir condition for a moving double layer was given by Ref.~\onlinecite{1992PhyS...45..391S} as $\Gamma_e\sqrt{m_e/m_i}=\Gamma_i-n_{\text{F}}U_{DL}$ where $n_{\text{F}}$ is the plasma density in the fireball and $U_{DL}$ is the double layer velocity in the lab frame. When there is an imbalance in the Langmuir condition there is a resulting velocity of the double layer.} 
and an increase of the fireball size. A restoring effect is produced by an increase in the plasma potential, which decreases $e\Delta\phi_{DL}$ as the fireball's electron collection area increases. Time resolved measurements of increases in plasma potential with fireball size have been experimentally observed for a fireball expanding to its equilibrium size just after onset, see Fig.~9 of Ref.~\onlinecite{1991JPhD...24.1789S}. 

The hysteresis can be understood by considering these elements of the model in conjunction with a consideration of global current balance of the entire plasma. Considering a positively biased electrode with sheath area $A_{\text{S}}$ in a plasma bounded by walls with area $A_{\text{W}}$, the plasma potential is determined by balancing the global electron and ion currents lost. Setting the electron current $I_e=e0.39n_ov_{T_e}[A_{\text{S}}+A_{\text{W}}\exp(-\frac{e\phi_p}{T_e})]$ equal to the ion current $I_i=e\Gamma_{i,B}A_{\text{W}}$ and solving for the plasma potential $\phi_p$ results in $\phi_{p,0}=-\frac{T_e}{e}\ln(\mu-A_{\text{S}}/A_{\text{W}})$, where $\mu=2.3\sqrt{m_e/m_i}$, $\Gamma_{i,B}=0.61n_o\sqrt{T_e/m_i}$ is the Bohm flux, and $0.39n_ov_{T_e}$ is the electron thermal flux\footnote{Although previous papers have demonstrated that the flux of electrons collected by an electron sheath is $n_ev_{T_e}$, the determination of the electron sheath edge density is an inherently 2D problem\cite{2016PhPl...23k3503H,2015PhPl...22l3520S}. Here, for simplicity, the flux is approximated to be the random flux. }. This will determine the initial plasma potential before fireball formation. The location of the initial plasma potential is marked with a red star on the experimentally measured electrode I-V and plasma potential-electrode bias curves in Fig.~1A and 1B. A corresponding drawing of a typical pre-onset sheath potential is shown in Fig.~1C. Increasing the electrode bias $\Delta \phi_c$ above $\phi_{p,0}$ will initiate fireball onset.

Once the fireball forms it will expand until the steady-state double layer potential $\Delta\phi_{DL}$ is reached. As the double layer expands its surface area $A_{\text{F}}$ increases, which results in an increase of plasma potential to compensate for the additional electron loss \footnote{ Using balance of global current loss with the steady-state Langmuir condition, the ratio of electron collection to ions sourced by the fireball is $\mathcal{O}(\sqrt{m_i/m_e})$.
Therefore, the increased electron loss area has the dominant effect on the plasma potential. }.
After expanding it reaches a value $\phi_{p1}$ given by
\begin{equation}\label{p1}
\phi_{p1}=-\frac{T_e}{e}\ln\bigg(\mu-\frac{A_{\text{F}}}{A_{\text{W}}}\bigg),
\end{equation}
the electrode bias still being fixed at $\phi_{E,1}=\phi_{p,0}+\Delta \phi_c$. 
This fireball configuration represented by the conditions $\phi_{E,1}$ and $\phi_{p,1}$ is marked with the red square in Fig.~1A and 1B, and a corresponding drawing of the fireball potential profile is shown in Fig.~1C. The relation between the fireball area and plasma potential given by Eq.~\ref{p1} is valid for any steady state fireball area $A_F$.

As long as $A_{\text{F}}/A_{\text{W}}<\mu$ for the initial fireball surface area prior to fireball expansion a stable fireball can exist. For the experimental parameters under consideration, the plasma potential is much greater than the wall potential when the fireball is present. Once the fireball forms, electrons are only lost to the fireball due to the large value of the bulk plasma potential and ions are only lost to the walls owing to the large double layer potential. The result is that the bulk plasma is in a state of global non-ambipolar flow. Due to this state, the value of the plasma potential is determined by the electrode potential. Furthermore, assuming the fireball plasma potential is near the electrode potential (discussed in the following paragraph), the fireball area is set by the current balance condition in Eq.~\ref{p1} and $\phi_{E,1}-\phi_{p1}\approx\Delta\phi_{DL}$. Therefore, the fireball area is determined in part by the chamber wall area $A_{\text{W}}$. This relation predicts that larger fireball areas are possible in larger plasma chambers.

Assuming that the plasma potential inside the fireball is near that of the electrode,
its value after steady-state is established is approximately $\phi_{E,1}-\Delta\phi_{DL}\approx\phi_{E,1}-\mathcal{E}_I/e$, see the blue line in Fig.~1B. Such an assumption can be justified for the experiments presented in this paper if the sheath between the fireball plasma and electrode is an ion sheath, which in a fireball typically has a small potential drop. From Ref.~\onlinecite{2017PhPl...24k3520S}, the condition for an ion sheath is $1+A_{\text{F}}/A_{\text{E}}<0.6\sqrt{T_{eI}m_i/2\pi T_{eC}m_e}$, where $T_{eI}/T_{eC}\sim1$ is the ratio of fireball electron temperatures for electrons from ionization and collection. Using the fireball surface area of $A_{\text{F}}=113 \ \textrm{cm}^2$ as measured in Fig. 2 (measurement to be described below) with the electrode area $A_{\text{E}}=2.84 \ \textrm{cm}^2$ this relation can be verified. 
 
Now consider the effect of decrementing the electrode bias to some value $\phi_{E,2}<\phi_{E,1}$ marked with the red circle in the panels of Fig.~1.
Immediately after decrementing the electrode bias, but before the bulk plasma potential has a chance to respond, the fireball will begin to contract due to an imbalance in the Langmuir condition resulting from insufficient ionization since the new double layer potential is less than the required value $\Delta\phi_{DL}$ given by Eq.~\ref{pdli}. The locking of the plasma potential to within $\sim\mathcal{E}_I/e$ of the electrode potential is mediated through changes in the fireball area. Changes in the plasma potential are caused by changes in the value of $A_F$ as described by Eq.~\ref{p1}.
Similar to the case described above, an underdensity in the fireball leads to the inward motion of the double layer. As the double layer contracts the plasma potential lowers to $\phi_{p,2}<\phi_{p,1}$. 
Once a sufficient fall in plasma potential is reached the required steady-state double layer potential $\Delta\phi_{DL}$ is attained once again. Now at $\phi_{E,2}$ a steady-state fireball can be maintained below the critical electrode bias given by $\phi_{E,1}$. The corresponding potential profile is indicated by the line marked with the circle in Fig.~1C.
This process can be repeated as long as the new, lower, electrode bias is not less than $\sim\Delta\phi_{DL}$ above the pre-fireball plasma potential.
 This description is consistent with the plasma potential being approximately $\phi_{E}-\mathcal{E}_I/e$ below the electrode bias, indicated by the blue line shown in Fig.~1B. The electrode bias at fireball collapse is slightly more than $\mathcal{E}_I/e$ above $\phi_{p,0}$ due to the order $T_e$ term in Eq. \ref{pdli}, which is typically a few eV. In Fig. 1 this is approximately 5V above $\mathcal{E}_I/e=16 V$, resulting in $\Delta\phi_{DL}\approx 21V$. Fireball collapse results from an inability of the plasma potential to decrease beyond the value given by Eq.~\ref{p1} when $A_F/A_W$ approaches the minimum ratio set by the electrode area $A_E/A_W$. In practice, the minimum fireball area just before extinction will be greater than this since the double layer surface area is always larger than $A_E$.

\begin{figure}
\includegraphics[scale=.55]{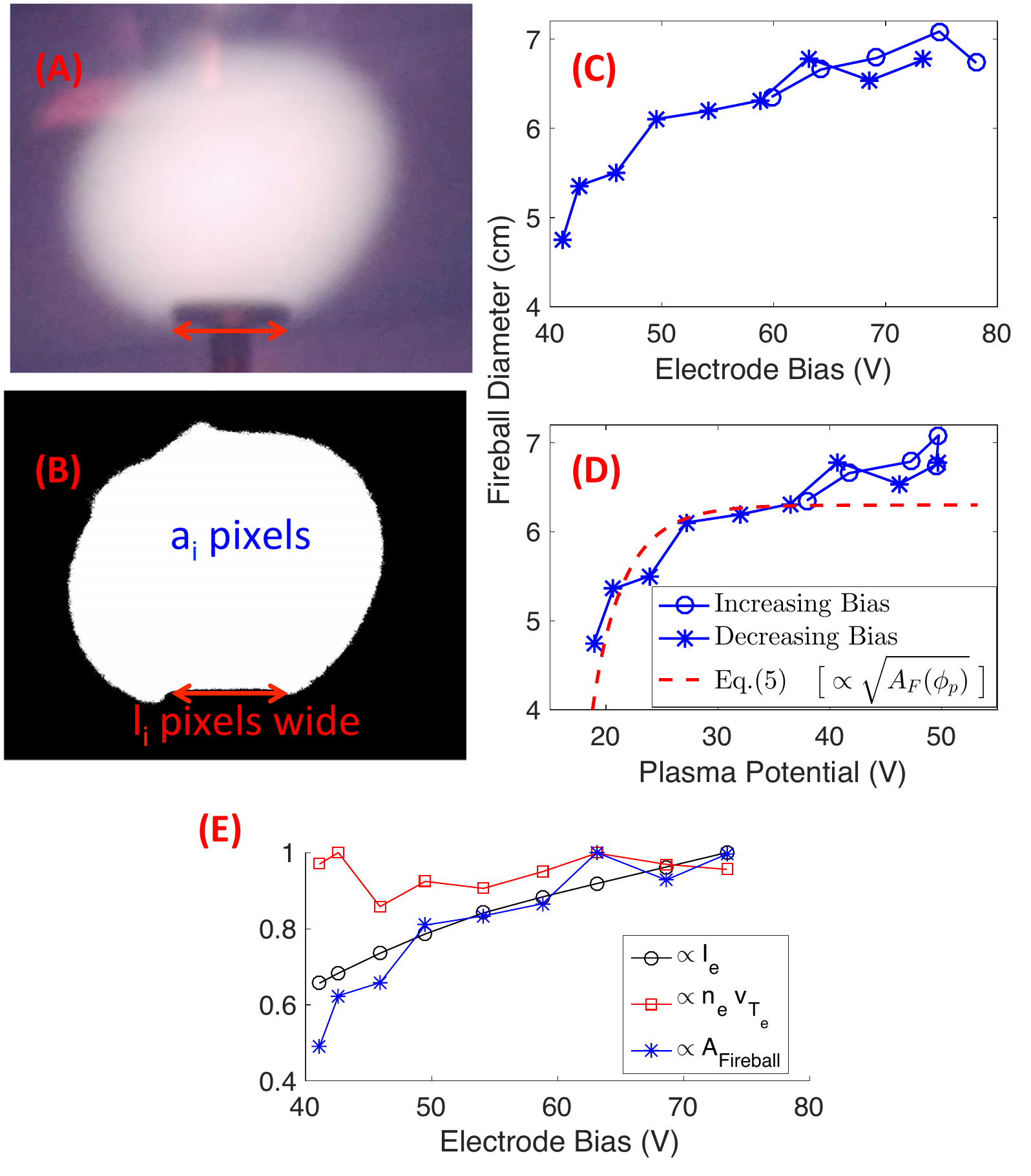}
\caption{A) An example of an image used to determine the fireball diameter. B) The same image converted to binary with a luminance threshold of 0.6. The fireball area $a_i$ and electrode length $d_i$ are marked in pixels. C) The fireball diameter as a function of electrode bias. D) The fireball diameter as a function of plasma potential. E) The scaling of electrode current, fireball area, and electron flux with electrode bias.}
\end{figure}

Additional experimental measurements were taken to verify the predicted relationship between fireball size, electrode bias, and plasma potential. These were conducted in a 1mTorr Ar filamentary discharge contained in a multi-dipole chamber which was described previously in Ref.~\onlinecite{2016PhPl...23k3503H}. Pre- and post-onset electron temperatures and densities were approximately 2-3.5eV and 1-2$\times10^9 cm^{-3}$. Fireball formation was initiated by biasing an insulated disk electrode with a conducting 1.9 cm diameter exposed face.
The plasma potential, electron density, and electron temperature were measured in the bulk plasma via a Langmuir probe sweep. The I-V and plasma potential-electrode bias plots for this experiment are shown in Fig.~1. To determine the size of the fireball as a function of electrode bias and plasma potential, a series of images were obtained along the hysteresis curve, see Fig. 2A. Each image was converted to a binary image with a luminance threshold set at 0.6 resulting in images like the one in Fig. 2B. The number of pixels in the white area $a_i$ were counted and used to define a characteristic diameter in pixels $d_i=\sqrt{4a_i/\pi}$. The characteristic fireball diameter corresponding to $d_i$ was determined by multiplying by the electrode diameter (1.9 cm) and dividing by its length in pixels $l_i$.
Figure 2C and 2D show that the fireball diameter as a function of electrode bias and plasma potential have a similar dependence. 
 
 Figure 2D corroborates the description of the hysteresis given above. A plot of a value proportional to $\sqrt{A_F}$ with plasma potential, calculated from Eq.~\ref{p1}, has a dependence which is in agreement with the experimental data. This demonstrates the coupling between the plasma potential and the fireball area. As the electrode bias is decremented below the critical bias, the fireball contracts and the plasma potential falls until a new equilibrium position where $e\Delta\phi_{DL}$ satisfies the relation in Eq.~\ref{pdli}. The constancy of the double layer potential is inferred from Fig.~1B. 
 
 The coupling between the plasma potential and fireball collection area is also apparent when evaluating the scaling of the electrode current. Figure 2E shows the electron current along with values of the electron flux inferred from experiments and the measured fireball surface area, all normalized to their maximum value. The electron flux density at the sheath edge of the fireball was calculated from experimental measurements using $n_e\sqrt{T_e/m_e}$. A comparison of these quantities during the decrementation of the electrode bias reveals that the electron flux is nearly constant, while the electrode current and the fireball area have similar scaling. Therefore, it is concluded that the fireball surface area is responsible for the change in current collection.

\begin{figure}
\includegraphics[scale=.55]{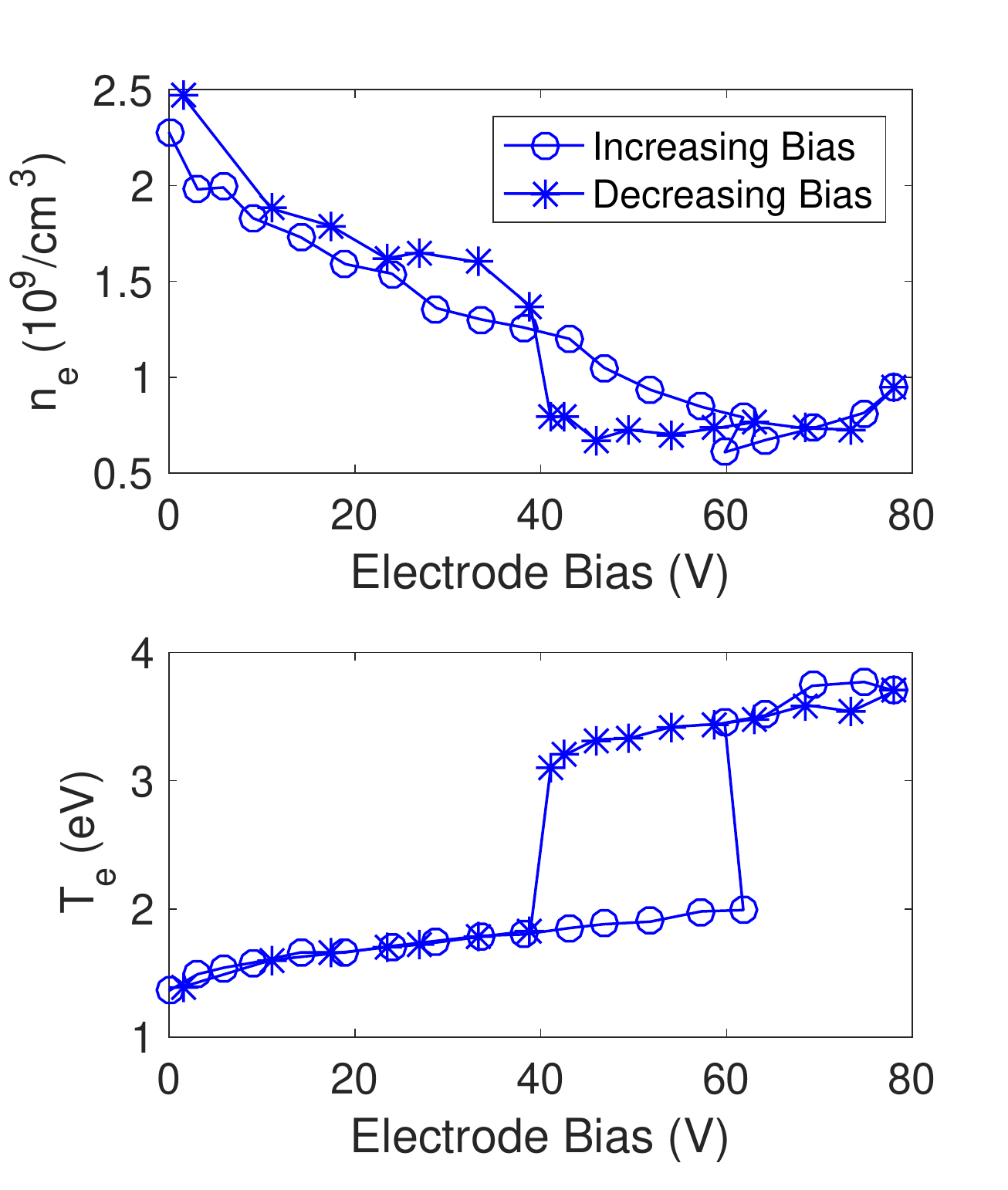}
\caption{The measured electron density and temperature as a function of electrode bias.}
\end{figure}

Another feature of the measurements is the change in temperature associated with the presence of the fireball, see Fig. 3. These experiments take place in a DC filamentary discharge plasma with a multidipole magnetic field configuration for plasma confinement \cite{2009PSST...18c5002B,2012PSST...21a5012S,2011PhPl...18f2113S,2011PhPl...18f2112S}. In these discharges, hot primary electrons from the filamentary cathode are accelerated by the large negative bias applied to the filament relative to the plasma potential. 
When the voltage between the plasma and chamber wall is small, only electrons with energy above the sheath voltage are collected. This results in a preferential loss of the most energetic electrons in the velocity distribution function to the wall. As the voltage between the plasma and the walls increases, fewer energetic electrons from the velocity distribution function are lost at this boundary. On the other hand, the electrode collects electrons indiscriminately. Thus, a greater electron temperature is achieved when electrons are only lost to the electrode. This behavior is typical of global non-ambipolar flow\cite{2007PhPl...14d2109B} and is concurrently present with the fireball as seen in Fig. 3. 

Figure 3 also indicates that the electron density decreases in the electron sheath regime, but is relatively constant when the fireball is present. The electron temperature on the other hand increases by $\sim1.5$eV when the fireball forms, indicating a state of global non-ambipolar flow. When the fireball is present, both quantities change relatively little, consistent with the near constant electron flux in Fig. 2E.

This paper presented a description and measurements of the hysteresis in fireball electrode I-V traces. 
The mechanism for the hysteresis was shown to be that the accelerating voltage needed to sustain the fireball through continued particle balance is significantly less than that needed to initiate its onset. After onset, the plasma potential locks to a value below that of the electrode by approximately the double layer potential $\Delta\phi_{DL}$ needed to maintain a steady-state fireball. 
Once a fireball is present, an increase or decrease of the electrode bias results in a corresponding increase or decrease in fireball size. This change in the fireball size either increases or decreases the plasma potential so that the steady state double layer potential can be maintained.
New experimental measurements demonstrate that the expansion and contraction of the fireball corresponds to the observed changes in the current collection and plasma potential. Although the discussion has focused on the I-V hysteresis as the electrode bias is varied, the same principals can be extended to the observed hysteresis with electrode bias and neutral pressure\cite{2009PSST...18c5002B,1991JPhD...24.1789S}.

This work was supported in part by US DOE Grant Award No. DE-SC0016473.

\bibliography{hysteresis_paper}

\end{document}